

\hsize=15.5 truecm
\vsize=22.2 truecm
\leftskip=1 truecm
\topskip=1 truecm
\splittopskip=1 truecm
\parskip=0 pt plus 1 pt
\baselineskip=19.9 pt
\def\re{$\rm I \kern-0.07cm R$}
\def\sq{\hfill \vbox {\hrule width3.2mm \hbox {\vrule height3mm \hskip 2.9mm
\vrule height3mm} \hrule width3.2mm }}
\def\a{\alpha}
\def\b{\beta}

\def\d{\partial}

\def\g{\gamma}

\def\k{\kappa}

\def\L{\cal L}

\def\tn{\cal N}
\def\ttn{{\cal P}}
\def\to{\cal O}

\def\tu{\cal U}

\def\s{\sigma}
\def\S{\Sigma}
\def\~{\tilde}
\def\ra{\rightarrow}
\def\(){(r,x^1,x^2)}
\def\('){(r',{x'}^1,{x'}^2)}
\def\x12{x^1,x^2}

\def\ne{neigh\-bor\-hood}

\vskip 3 cm
\centerline{\bf GLOBAL EXTENSIONS OF SPACETIMES}
\centerline{\bf DESCRIBING ASYMPTOTIC FINAL STATES OF BLACK HOLES}
\bigskip
\centerline{{\sl Istv\'an R\'acz}}
\medskip
\centerline{MTA KFKI Research Institute for Particle and
Nuclear Physics}

\centerline{H-1525 Budapest 114, P.O.B. 49, Hungary }

\centerline{and}

\centerline{\sl Robert M. Wald} \medskip
\centerline{Enrico Fermi Institute and Department of Physics}

\centerline{University of Chicago, Chicago IL 60637, USA}
\bigskip
\bigskip
\centerline{\sl Abstract }
\par

We consider a globally hyperbolic,
stationary spacetime containing a black hole but no white hole.
We assume, further, that the event horizon, $\tn$, of the
black hole is a Killing horizon with compact cross-sections. We prove that if
surface gravity is non-zero constant throughout the horizon
one  can {\it  globally} extend such a spacetime
so that the image of $\cal N$ is a proper subset of a
regular bifurcate Killing horizon in the  enlarged
spacetime. The necessary and sufficient conditions are given for
the extendibility of matter fields to the enlarged spacetime. These
conditions are automatically satisfied if
the spacetime is static (and, hence ``$t$"-reflection symmetric)
or stationary-axisymmetric with ``$t-\phi$"
reflection isometry and the matter fields respect the reflection isometry.
In addition, we prove that a necessary and sufficient condition for the
constancy of the surface gravity on a Killing horizon is that the exterior
derivative of the twist of the horizon Killing field vanish on the horizon.
As a corollary of this, we recover a result of Carter
that constancy of surface gravity holds for any black hole
which is static or stationary-axisymmetric with the
``$t-\phi$" reflection isometry.
No use of Einstein's equation is made in obtaining any of the above results.
Taken together, these results support the view that
any spacetime representing the asymptotic final state of a black hole
formed by gravitational collapse may be assumed to possess a bifurcate
Killing horizon or a Killing horizon with vanishing surface gravity.

\medskip
\parindent 20pt
PACS number: 04.20 Cv.
\vfill\eject
\bigskip
\parindent 0 pt
{\bf 1. Introduction}
\parindent 20 pt
\medskip

It is of considerable interest
to determine of the possible asymptotic
final states of the gravitational collapse
of an isolated body. The ``cosmic censor hypothesis" conjectures
that gravitational collapse always produces a black hole, in such a way
that a neighborhood containing the exterior region together with
the event horizon of the
black hole is globally hyperbolic (see, e.g., the discussion in [1]).
In addition, it is widely
expected that the asymptotic
final state of such a collapse can be represented by a stationary spacetime
(i.e., a spacetime possessing a one-parameter group of isometries whose
orbits are timelike near infinity).
As we shall review in the next section, arguments of
Hawking [2,3] and Carter [4,5] show that, in a wide variety of
circumstances, the event horizon of a stationary black
hole must be a {\it Killing horizon}, by which we mean
a null hypersurface (i.e., an embedded submanifold of co-dimension one)
whose
generators coincide with the orbits of a one-parameter group of isometries.
Therefore, the study of globally hyperbolic spacetimes with
a Killing horizon is of considerable
importance with regard to the classification of the possible final states
of gravitational collapse.

In a previous paper [6], we analyzed the extendibility of spacetimes with a
one-parameter  group  of  isometries  possessing a Killing horizon,
$\cal N$, such that the generators on $\cal N$ are diffeomorphic to \re
\ and  that $\cal  N$ admits  a smooth  cross-section.
We showed that whenever the gradient,
$\nabla_a\k$, of the  surface
gravity, $\k$,  of $\cal  N$ is non-zero
on a generator of $\cal  N$,  then that
generator terminates in a parallelly propagated curvature
singularity, so
no extension can exist where $\cal N$ comprises a portion
of a regular, bifurcate Killing horizon (see also [7,8]). On the other hand, if
$\k$ is  constant and  nonvanishing on $\cal N$,
we proved that a neighborhood, $\cal U$,
of $\cal N$ always  can be extended so that $\cal N$ comprises  a
portion of a bifurcate Killing horizon.
However, this analysis did not determine the conditions under which
such an extension could be performed {\it globally}
-- i.e., such that not merely a  \ne, $\cal U$, of the horizon  but
the entire spacetime can be imbedded into a larger spacetime which
possesses a bifurcate  Killing horizon. The existence of a global extension
would be needed to argue that, without loss of generality, one can assume
that $\cal  N$ comprises a portion of a bifurcate Killing horizon.

One might expect that the condition of global hyperbolicity of the
original spacetime would
suffice, by itself, to ensure the existence of the desired global extension.
However, the following example shows that this is not the case,
i.e., there exist globally  hyperbolic
spacetimes possessing a Killing horizon with
constant but non-zero surface gravity which can not be extended
so that the Killing horizon is a portion of a regular, bifurcate horizon.

 \parindent 0 pt {\it Example 1.:} \parindent 20 pt Start with the
3-dimensional  Minkowski   spacetime,  (\re$^3,\eta_{ab})$,   with
Cartesian coordinates, $t,x,y$, and consider the boost  isometries
about the origin
in the $t - x$  plane.  Now consider the spacelike  surface, $\cal
S$,  comprised  by  all  Killing  orbits which intersect the line,
$x=0$, $t=-\vert  y\vert/2$.  Let  $\Omega$ be  a smooth  function
which is defined on the chronological future, $I^+[{\cal S}]$,  of
$\cal S$, such that it takes the value $1$ whenever $t \geq -\vert
x\vert$, it  is ``boost-invariant",  and for  which the  curvature
scalar of  the spacetime,  $(I^+[{\cal S}],  \Omega^2 \eta_{ab})$,
blows up everywhere  `on' ${\cal  S}\setminus\{(t,x,y)\vert
t^2-x^2=0\ {\rm and}\ y=0\}$.  An example
of such a function, $\Omega$, is
$$\Omega(t,x,y)=\cases{f(2(t^2-x^2)^{1\over  2}/\vert  y\vert),&if
$0\leq  2(t^2-x^2)^{1\over   2}/\vert  y\vert   <1$,  $t<0$,   and
$y\not=0$;\cr          1,&otherwise,\cr}\eqno(1)$$           where
$f(z)=1-e^{(1-z^{-2})}$, which is well defined for $0\leq z<1$.
The spacetime, $(I^+[{\cal S}], \Omega^2 \eta_{ab})$, is the
desired globally  hyperbolic spacetime.   The hypersurface $t = x > 0$
(as well as the hypersurface $t = -x > 0$) is Killing horizon,
on which the null generator at $y = 0$
is geodesically incomplete generator.  An open
neighborhood of this Killing horizon in the spacetime
$(I^+[{\cal S}], \Omega^2 \eta_{ab})$ is isometric to an open
neighborhood of the corresponding Killing horizon in the
spacetime, $(I^+[{\cal S}],\eta_{ab})$, so, clearly, the the spacetime
$(I^+[{\cal S}], \Omega^2 \eta_{ab})$ is locally
extendible to a spacetime
with a  bifurcate horizon, in accord with the results of [6].
However, it also is clear that there is no global
extension of $(I^+[{\cal S}], \Omega^2 \eta_{ab})$
to a spacetime with a  bifurcate horizon,
since there is no way to ``put back" the origin
so that the generator at $y = 0$ will extend through a bifurcation surface.
Thus,   $(I^+[{\cal   S}],   \Omega^2   \eta_{ab})$,   is  a
globally  hyperbolic spacetime possessing
a non-bifurcate Killing horizon with constant, nonzero surface gravity,
which cannot be globally extended to a spacetime possessing
a bifurcate Killing horizon. \sq

The difficulty occurring in the above example
can be seen to arise from the fact that
a portion of the ``bottom wedge" (as well as the
entire ``left wedge") is already present in the original spacetime, and
they ``get in the way" of any global extension of the
portion, $t = x > 0$, of the horizon.
However, in a spacetime representing the asymptotic final
state of a stationary black hole, the ``left" and ``bottom" wedges would
correspond to a white hole region of the spacetime. Such a white hole
would not occur in the spacetime describing an actual, physically realistic
gravitational collapse, but rather would
correspond to a physically irrelevant,
``early time" region of the spacetime describing the black hole final
state.

In this paper, we will restrict attention to globally hyperbolic,
stationary spacetimes (with one asymptotic end) which contain a black
hole but no white hole. It should be noted that given any globally
hyperbolic, stationary
spacetime which contains a white hole, the
sub-spacetime comprised by $I^+[{\cal I}^-]$
also is stationary and globally hyperbolic but does not
contain a white hole. Thus, our assumption that no white hole is present
in the spacetime actually
involves no loss of generality
with regard to the class of spacetimes we consider, but rather
merely eliminates a physically irrelevant
region of these spacetimes. We shall assume, in addition,
that  the event horizon,
$\tn$, of the black hole is a Killing horizon and that $\tn$ has compact
cross-sections. Our main result is the following: If (and only if) the
surface gravity is a non-zero constant on $\tn$,
one can {\it globally}
extend the a spacetime so that the image of $\cal N$ is a proper subset of a
regular bifurcate Killing horizon in the enlarged spacetime.

In the next section, some properties of stationary black holes will be
reviewed. In addition, we shall give a simple proof that the
necessary and sufficient condition of the constancy of $\k$ is
the vanishing  on the horizon
of the exterior derivative, $\nabla_{[a} \omega_{b]}$,
of the twist, $\omega_a\ ( \equiv
\epsilon_{abcd}\xi^b\nabla^c\xi^d$), of the horizon Killing field $\xi^a$.
A corollary of this result yields a result of Carter [5] establishing that if
the black hole is static or stationary-axisymmetric with the
``$t-\phi$" reflection isometry, then the surface gravity
must be constant over the horizon. Thus, the ``zeroth law" of black
hole mechanics holds in this context without the imposition of
any field equations.

Some properties of the space of Killing orbits are established in section 3,
in preparation for the proof of our main result on global extensions given
in section 4. The extendibility of matter fields is analyzed in section 5.

Finally, we emphasize that, as in our previous paper [6],
no use of Einstein's (or any other) field equation
is   made   anywhere   in   our analysis.
Furthermore,  although  for  definiteness  we  treat the case of a
4-dimensional   spacetime,   all   of   our   results   generalize
straightforwardly to any spacetime dimension $n\geq 2$.
Similarly, although for simplicity, we shall assume that the black hole
event horizon, $\tn$, is connected
our results generalize straightforwardly to
the case where $\tn$ possesses disconnected components.

\bigskip
\parindent 0 pt
{\bf 2. Stationary black holes and the Zeroth Law}
\parindent 20 pt
\medskip

In this section,
we shall give a simple proof of a generalization of a result of
Carter on the constancy of the surface gravity of static and
stationary-axisymmetric black holes, and we shall specify the
precise class of spacetimes we shall consider in the following sections.

We begin by briefly recalling the arguments
that the event horizon of a stationary black hole must be a
Killing horizon. Actually, there are two complementary approaches
which lead to this conclusion. The argument of
Hawking [2,3] assumes that
Einstein's equation holds with matter satisfying suitable hyperbolic
equations and the dominant energy condition. Moreover, it
assumes that the spacetime is analytic. By consideration of the null
initial value formulation (with one of the two intersecting
null hypersurfaces taken to
be the event horizon of the black hole), it is then shown that the initial
data must be invariant under
the action of a one parameter group whose orbits on the event horizon
coincide with its null geodesic generators. By uniqueness of the Cauchy
evolution together with analyticity, it then follows that the spacetime
possesses a one-parameter group of isometries, with Killing field normal
to the event horizon, i.e., that the event horizon is a Killing horizon.
Note that
if the stationary Killing field fails to be normal to the event horizon, this
argument shows that there must exist an additional Killing field. Further
arguments establish that, in this case,
a linear combination of the stationary Killing field and the Killing field
orthogonal to the horizon must have closed orbits, so that the spacetime
is axisymmetric as well as stationary.

The argument of Carter [4,5] {\it assumes} that the black hole is
either static or is stationary-axisymmetric with a
``$t-\phi$" reflection isometry. In the static case,
it then is shown that the black hole event horizon
must coincide with (a portion of) the
``staticity limit", defined to be the boundary of the
region where the static Killing field is timelike. Similarly, in the
stationary-axisymmetric case, the black hole event horizon
must coincide with (a portion of) the
``circularity limit", defined to be the boundary of the
region where there exists a linear combination of the
stationary and axial Killing fields which is timelike. In either case,
it then is shown that the event horizon must be a Killing horizon. In
the static case, the static Killing field itself must be orthogonal to the
event horizon. In the stationary-axisymmetric case, some constant linear
combination of the stationary and axial Killing fields must be orthogonal
to the horizon. Note that
although this argument makes considerably stronger assumptions
about the symmetries of the black hole spacetime compared with
the argument of the previous paragraph, Einstein's
equation is not used, so the argument is applicable to much more
general theories of gravity.

Next, we recall the surface gravity, $\kappa$,
of a Killing horizon, $\tn$, with Killing field $\xi^a$ is defined
on $\tn$ by the equation
$${1\over 2}\nabla^a(\xi^b\xi_b)= -\kappa\ \xi^a.\eqno(1)$$
It follows immediately that $\kappa$ is constant along the orbits of
$\xi^a$. Our main new result of this section is the following:

\medskip\parindent 0pt
\leftskip=3.3 truecm

\item{\it Theorem 2.1:} Let $\tn$ be a connected
Killing horizon, with Killing field
$\xi^a$. Then the surface gravity, $\kappa$, is constant on $\tn$
if and only if the exterior
derivative of the twist form field, $\omega_a$, is zero on the horizon, i.e.,
$$\nabla_{[a}\omega_{b]}\vert_{_{\tn}}=0,\eqno(2)$$
where $\omega_a$ is defined by
$\omega_a =\epsilon_{abcd}\xi^b \nabla^c \xi^d$.

\leftskip=1 truecm

{\it Proof:} By a standard relation satisfied by the surface gravity (see,
e.g.,
eq. (12.5.30) of [1]), we have
$$\xi_{[a}\nabla_{b]}\kappa= -\xi_{[a}{R_{b]}}^e\xi_e.\eqno(3)$$
On the other hand, by a standard identity satisfied by the twist (see, e.g.,
eq. (7.1.15) of [1]), we have
$$\nabla_{[a}\omega_{b]}=-\epsilon_{abcd}\xi^{[c}{R^{d]}}_e\xi^e.
\eqno(4)$$
Thus, on the horizon, we have
$$\xi_{[a}\nabla_{b]}\kappa= - {1 \over 4}
\epsilon_{abcd}\nabla^{[c} \omega^{d]}\eqno(5)$$
from which the theorem follows immediately. \sq

\leftskip=1 truecm
\parindent 20 pt

This theorem has, as a consequence, the following corollary,
which expresses the essential content of a result previously obtained
by Carter (see theorem 8 of [5]; see also Heusler [9]):

\medskip\parindent 0pt
\leftskip=3.3 truecm

\item{\it Corollary 2.2:} Let $\tn$ be a connected
Killing horizon, with Killing field
$\xi^a$. Then, (i) If $\xi^a$
is hypersurface orthogonal, then $\kappa$ must be constant throughout the
horizon. In particular, $\kappa$ is constant on the
horizon of any static black hole.
(ii) If there exists a
Killing field, $\psi^a$, on the spacetime which is linearly independent of
$\xi^a$, commutes with
$\xi^a$, and on the horizon satisfies $\nabla_a(\psi^b\omega_b) = 0$,
then
$\kappa$ is constant on $\tn$. In particular, $\kappa$ is constant on the
horizon of any
stationary-axisymmetric black hole possessing the ``$t-\phi$" reflection
isometry.

\leftskip=1 truecm

{\it Proof:} The first claim of (i) is trivial, since $\omega_a = 0$
everywhere in the hypersurface orthogonal case. The second claim
of (i) follows from the fact [5] (reviewed above) that the event horizon
of a static black hole is a Killing horizon, with $\xi^a$ equal to the static
(and, thus, hypersurface orthogonal) Killing field.

\parindent 20pt

To prove the first
claim of (ii), we assume the contrary. Then there exists an open
subset, ${\cal O}$, of
$\tn$ such that in ${\cal O}$ we have $\xi_{[a}\nabla_{b]}\kappa \neq 0$.
Clearly then,
without loss of generality, we may assume that $\kappa\not=0$ in $\cal O$.
Since $\psi^a$ is a Killing field which commutes with $\xi^a$, we have
$$0 = {\L}_\psi (\xi^a \xi_a) = - 2 \kappa \psi^a \xi_a,\eqno(6)$$
so $\psi^a$ is tangential to $\tn$ in ${\cal O}$. Now, if $\psi^a$ were
proportional to $\xi^a$ in ${\cal O}$ -- i.e., if $\psi^a = f \xi^a$
in ${\cal O}$ --
then the commutativity of $\psi^a$ and $\xi^a$ implies that
$\xi^a \nabla_a f = 0$ in ${\cal O}$. On the other hand, since
$\psi^a \psi_a = 0$ in ${\cal O}$, it follows that ${\cal O}$ also is
(a portion of) a Killing horizon with respect to $\psi^a$, with surface
gravity with respect to $\psi^a$ given by $\tilde{\kappa} = f \kappa$.
Application of eq. (3) to both $\kappa$ and $\tilde{\kappa}$ then
implies that $\xi_{[a} \nabla_{b]} f = 0$. Thus, we have $\nabla_a f = 0$
in ${\cal O}$, which, in turn, implies that $\psi^a$ and $\xi^a$ are linearly
dependent as Killing fields, contrary to our hypothesis. This shows that
there must exist a point $p \in {\cal O}$ where $\psi^a$ is strictly
spacelike. Now, since $\psi^a$ is a Killing field which commutes
with $\xi^a$, it follows immediately from eq. (1) that
$\psi^a \nabla_a \kappa = 0$ at $p$. On the other hand, from eq. (5)
we obtain
$$\epsilon^{efab} \psi_f \xi_a \nabla_b \kappa =
\psi_f  \nabla^{[e} \omega^{f]}\eqno(7)$$
But, since $\omega_a$ vanishes on $\tn$ (since $\xi^a$ is hypersurface
orthogonal on $\tn$) and since ${\L}_\psi \omega_a = 0$ everywhere
(since $\psi^a$ is a Killing field which commutes
with $\xi^a$), the right side of eq. (7) can be rewritten as
$$\psi_f  \nabla^{[e} \omega^{f]} = {1 \over 2} \nabla^e (\psi_f
\omega^f)\eqno(8)$$
which vanishes by hypothesis. Thus, both
$\epsilon^{efab} \psi_f \xi_a \nabla_b \kappa$ and
$\psi^a \nabla_a \kappa$ vanish at $p$, which implies that
$\xi_{[a}\nabla_{b]}\kappa = 0$ at $p$, in contradiction to our
hypothesis. This establishes the first claim of (ii).

Finally, the second claim of (ii) follows from the fact [5] (reviewed above)
that the event horizon of a stationary-axisymmetric black hole with
``$t-\phi$" reflection symmetry is a Killing horizon with respect to some
constant linear combination, $\xi^a$, of the stationary and axial Killing
fields. If we choose $\psi^a$ to be the axial Killing field, then the Frobenius
integrability condition for the ``$t-\phi$" reflection
symmetry requires that $\psi^a \omega_a = 0$ throughout the spacetime,
so the hypothesis of the first claim of (ii) holds. \sq

We emphasize that in the derivation of the above results, no restriction has
been imposed on the causality structure of the spacetime or on the
topological properties of the Killing horizon (except for our simplifying
assumption that $\tn$ is
connected). In particular, these results are valid for any
metric theory of
gravity, including theories with torsion (with the understanding that
in our formulas, $\nabla_a$ always denotes the torsion free, metric
compatible derivative operator).

\leftskip=1 truecm
\parindent 20 pt

We conclude this section by giving a mathematically
precise specification of the
class of spacetimes, $(M,g_{ab})$, which will be
considered in the remainder of this paper.
We require that $(M,g_{ab})$ be a smooth, globally hyperbolic
spacetime, possessing a smooth spacelike Cauchy
surface $\Sigma$ which can be expressed in the form
$\Sigma=\Sigma_{end}\cup\Sigma '$ so that $(M,g_{ab})$ is
$(k,\a)$-asymptotically stationary with respect to the single
asymptotically flat ``end"
$\S_{end}$, as specified in definition 2.1 of [10]. However, it should
be noted that the precise
asymptotic flatness conditions on $\S_{end}$
given in that definition will not be of great
importance here, and could be significantly weakened or modified.
We denote by $\phi_u$ the one-parameter group of isometries whose
orbits are timelike on $\S_{end}$, and denote its associated Killing field
by $t^a$. We define $M_{end}$ to be the
orbit of $\S_{end}$ under these isometries,
$$M_{end}=\phi\{\Sigma_{end}\}.\eqno(9)$$
The {\it black hole region}, ${\cal B}$, of $M$ is defined to be the complement
of $I^-[M_{end}]$, and the {\it white hole region}, ${\cal W}$, is
defined to be the complement of $I^+[M_{end}]$.
As discussed in the introduction, we shall assume that
${\cal B} \neq \emptyset$ but ${\cal W} = \emptyset$, so that
$$M = I^+[M_{end}].\eqno(10)$$
The {\it future event
horizon} of the black hole is defined by
$${\tn}=\partial I^-[M_{end}],\eqno(11)$$
As discussed above, we shall assume that ${\tn}$ is a Killing horizon, i.e.,
that there exists a one-parameter group of isometries $\chi_u$
(possibly different from $\phi_u$) with Killing field $\xi^a$ normal to
${\tn}$. We shall assume that the intersection of $\tn$ with $\Sigma$ is
compact. In addition,
for simplicity, we assume that ${\tn}$ is connected. By reversing
the sign of $\xi^a$ if necessary, we may assume that the orbits of $\xi^a$
are future directed on ${\tn}$.
If $\xi^a \neq t^a$, it follows from our asymptotic
conditions that a linear combination of $\xi^a$ and $t^a$ has closed
orbits, i.e., that the spacetime is stationary and axisymmetric. In
particular -- whether or not $\xi^a = t^a$ -- it follows that $(M,g_{ab})$ is
$(k,\a)$-asymptotically stationary-rotating with respect to $\xi^a$ (see
the definition of Appendix A of [10]); in particular, the orbits of $\xi^a$
in $M_{end}$ are future
oriented, i.e., for each $p\in M_{end}$ there exists
an increasing sequence $\{u_i\}$
such that $\chi_{u_{i+1}}(p)\in
I^+\bigl(\chi_{u_i}(p)\bigr)$. We note that
our results could be generalized
straightforwardly to spacetimes which are merely stationary-rotating
with respect to the Killing field, $\xi^a$, which is normal to the horizon,
without the need to assume the existence of the asymptotically
stationary Killing field $t^a$.

\bigskip
\parindent 0 pt
{\bf 3. Space of Killing orbits }
\parindent 20 pt
\medskip

In this section we shall consider the class of spacetimes, $(M,g_{ab})$,
specified at the end of the previous section and we will establish
some properties of the space of  Killing orbits
which will be needed to perform our global extension.
Our first result is the following:

\medskip\parindent 0pt
\leftskip=3 truecm

\item{\it Lemma 3.1:} Let $(M,g_{ab})$ be a spacetime in the class
specified at the end of the previous section. Then
no zero of $\xi^a$ exists in $M$.

\leftskip=1 truecm
\parindent 0 pt

{\it Proof:} Suppose, on the contrary, that $\xi^a\vert_p=0$ for some $p\in
M$. Since $p$ is invariant under the isometries, it follows from lemma 3.1
of [10] that either $I^-(p) \cap M_{end} = \emptyset$ or
$I^-(p) \supset M_{end}$. The first possibility is excluded by our
assumption that $M = I^+[M_{end}]$. However, the second possibility
would imply that $I^-(p) \supset \S_{end}$, which contradicts the fact that
$J^-(p)$ has to intersect any Cauchy surface $\Sigma$ in a compact set. \sq

\medskip
\parindent 20 pt

Since, in particular, the horizon, $\tn$, contains  no
fixed  point  of  $\chi_u$, any smooth spacelike
Cauchy surface, ${\cal C}$, intersects $\tn$ in
a global cross-section, $\sigma$.
This fact combined with the fact that for globally  hyperbolic spacetimes
the Killing orbits lying  on $\tn$ are  diffeomorphic to \re\  implies
that condition 2.1 of Ref. [6] is automatically satisfied for the class of
spacetimes considered here.

Let  ${\cal  C}$ be an arbitrary smooth, spacelike
Cauchy  surface, and  let $\s$ denote
the intersection,  $\tn\cap\ {\cal C}$, of
the Killing horizon and the Cauchy surface.
Then $\s$ is a smooth, embedded submanifold of ${\cal  C}$, which, by our
assumption above, must be compact. There are  two families of
null geodesics the  members of which  are orthogonal to  $\sigma$.
One of these families  generates the  Killing horizon,
$\tn$, while the members of  the other congruence determine --  at
least in a  sufficiently small \ne\  of $\sigma$ --  a smooth null
hypersurface, ${P}$. The next lemma shows that in a sufficiently
small neighborhood of $\s$, ${P}$ comprises a portion of boundary
of the future and past of $\s$.

\medskip\parindent 0pt
\leftskip=3 truecm

\item{\it Lemma 3.2:} There exists an open neighborhood,
${\cal U}$, of $\s$ in
$M$ such that $[{P} \cap J^+[\s] \cap {\cal U}] \subset \partial J^+[\s]$,
and, similarly, $[{P} \cap J^-[\s] \cap {\cal U}] \subset \partial J^-[\s]$

\leftskip=1 truecm

 {\it Proof:} It suffices to prove the first claim, since the second claim
follows by interchanging futures and pasts.
Consider the map from the normal bundle of $\s$ into $M$
which assigns to each $(s, n^a)$ in the normal bundle
(so that $s \in \s$ and $n^a$ is in the tangent
space at $s$ and normal to $\s$) the point in $M$ lying at a unit affine
parameter along the geodesic determined by $(s, n^a)$. Then this map is
smooth, and, by the implicit function theorem, there exists a neighborhood
${\cal U}_1$ of $\s$ such that this map is one-to-one and onto. Let
${\cal U}_2$ be a causal normal neighborhood of ${\cal C}$ (see lemma
2.2 of [8]) and let ${\cal U} = {\cal U}_1 \cap {\cal U}_2$. Let
$p \in {P} \cap J^+[\s] \cap {\cal U}$.
For $s \in J^-(p) \cap \s$ define $\tau_p(s)$ to be the length
(defined to be $\ge 0$) of the unique
causal geodesic connecting $p$ with $s$; for
$s \in \s$ but $s \notin J^-(p)$, define $\tau_p(s)$ to be $0$. Then
$\tau_p(s)$ is a continuous function of $s$. Since $J^-(p)$ is closed
in a globally hyperbolic spacetime, it follows that
$J^-(p) \cap \s$ is compact and non-empty, so
$\tau_p(s)$ must achieve a maximum at a point $s_0 \in \s$. It follows
that $p$ must be connected to $s_0$ by a causal geodesic which is
orthogonal to $\s$. However, since $p \in {\cal U}_1$, there is a unique
geodesic from $p$ to $\s$ which is orthogonal to $\s$, and since
$p \in {P}$, this geodesic is a null geodesic. Hence, $\tau_p(s) = 0$
for all $s \in \s$, i.e., $p$ cannot be connected to any point of $\s$ by
a timelike geodesic. It follows that $p \in \partial J^+[\s]$, as we desired
to show. \sq

\bigskip\parindent 20pt

We now define $\ttn = {P} \cap {\cal U}$.
The next lemma
shows  that no  timelike  curve can start on
$\ttn$   and   return to
${\ttn}$.   Similarly,  Killing   orbits  are  shown   to
intersect $\ttn$ precisely once.

\medskip\parindent 0pt
\leftskip=3.3 truecm

\item{\it Lemma 3.3: {\rm a)}}
${\ttn}$ is an achronal hypersurface.

 \item{\phantom{\it  Lemma  2.1:} b)}
Any   Killing   orbit   starting   on
$\ttn$ never intersects
$\ttn$ again.

\leftskip=1 truecm

 {\it Proof:} a)\ \ Let $p,q \in \ttn$. We wish to show that
there does not  exist  a
timelike  curve,  $\g$, from  $p$ to  $q$. If $p,q \in \partial J^+[\s]$
or if $p,q \in \partial J^-[\s]$, the result follows
immediately from the previous
lemma together with the achronality of causal
boundaries. Thus, we need only consider the case
$p \in \partial J^-[\s] \cap \ttn$ and  $q \in \partial J^+[\s] \cap \ttn$.
In this case, any timelike curve
$\g$ from $p$ to $q$ must intersect ${\cal C}$ at a point $r$.
However, if $r \in {\cal B}$ we would obtain a contradiction
with $p \in \partial J^-[\s] \cap \ttn$, since
$[\partial J^-[\s] \cap \ttn] \subset \partial J^-[{\cal C} \cap {\cal B}]$.
Similarly, if $r \notin {\cal B}$, we would contradict
$q \in \partial J^+[\s] \cap \ttn$.

\parindent 20pt

 b)\   \   We wish to show that for any $p \in {\ttn}$ there does not exist
a $u > 0$ such that $\chi_u(p) \in {\ttn}$. If $p \in \s$, the Killing orbit
through $p$ is a (future-directed) null geodesic generator of $\tn$, and
$\chi_u(p) \in I^+[{\cal C}]$ for all $u > 0$. Since ${\tn} \cap {\ttn}= \s$,
it follows that $\chi_u(p) \notin {\ttn}$. On the other hand, if
$p \in J^+[\s] \setminus \s$, then there exists a future-directed null
geodesic $\lambda$ from a point $s \in \s$ to $p$. Applying the
isometry $\chi_u$ to this statement, we find that there is a
future-directed null geodesic joining $\chi_u(s)$ to $\chi_u(p)$. Hence,
$s$ may be joined to $\chi_u(p)$ by a (future-directed) broken
null geodesic. It follows that $\chi_u(p) \in I^+[\s]$ and, hence, by part
(a) of this lemma we have $\chi_u(p) \notin {\ttn}$. The remaining
case $p \in J^-[\s] \setminus \s$ follows similarly. \sq

\bigskip

The above lemma shows that the isometry invariant
open \ne, $\to$$_{\tn} = \chi\{\ttn\}$, of the horizon, $\tn$
possesses the structure of a trivial
principal fibre bundle, with structure group
\re.  Nevertheless, it remains possible that the Killing orbits in
$\to$$_{\tn}$ could come arbitrarily close to other Killing orbits
in $M$ which do not intersect $\partial \ttn$.
(Note that if that occurred, the space
of Killing orbits in $M$ would fail to have Hausdorff topology.) If this
were to happen, these additional Killing orbits could ``get in the way"
of an attempted
global extension of $M$. The following provides a relevant example of a
globally hyperbolic spacetime possessing a Killing horizon where
Killing orbits in
$\to$$_{\tn}$ come arbitrarily close to Killing orbits which do not
intersect the closure of $\ttn$.

 \parindent 0 pt {\it Example 2:} \parindent 20 pt Let $(M, g_{ab})$
be  Minkowski  spacetime  with the  ``bottom wedge"
$t \leq - |x|$ removed.
The resulting  spacetime is globally hyperbolic and possesses a
one parameter group of isometries generated by the boost Killing
field $\xi = t {\partial \over \partial x} + x {\partial \over \partial t}$,
which is nowhere vanishing on $M$. Furthermore, the hypersurface
$t = x$ is a Killing horizon with respect
to $\xi^a$. Nevertheless, for any choice of $\ttn$, the
closure of the set $\to$$_{\tn} = \chi\{\ttn\}$ includes all of the Killing
orbits with $t = - x$.

\medskip

The next lemma shows that the type of behavior exhibited in the
above example cannot occur for the class of spacetimes considered
here.

\leftskip=1 truecm

\medskip\parindent 0pt
\leftskip=3 truecm

\item{\it Lemma 3.4:}
Let $\tn$, $\s$ and $\ttn$ be defined as above. Let ${\ttn}^*$ to be
an open subset of ${\ttn}$ with compact closure in $\ttn$.  Then the
boundary (in $M$) of the set $\chi\{{\ttn}^*\}$ is generated by orbits
meeting ${\ttn}$ at the boundary of ${\ttn}^*$.

\medskip\leftskip=1 truecm

{\it Proof:} Suppose on the contrary that there exists $q\in\partial
\bigl[ \chi \{{\ttn}^*\}\bigr]\setminus \chi\{\partial
{\ttn}^*\}$ in $M$. The statement that $q\in\partial [\chi \{{\ttn}^*\}]$
implies that there exists a sequence $\{p_i\}$ in $\ttn$ and a sequence
of real numbers $\{u_i\}$ such that the sequence
$\{q_i = \chi_{u_i} (p_i) \}$ converges to $q$. Since ${\ttn}^*$ has compact
closure in $\ttn$ there must exist a subsequence of the $\{p_i\}$ converging
to a point $p$ in $\overline {{\cal P}^*} \subset {\ttn}$.
Passing to this subsequence, we note that if the sequence $\{u_i\}$
had an accumulation point, $u$, then we would have $q = \chi_u(p)$,
in contradiction to our hypothesis. It follows that we must have either
$u_i\rightarrow - \infty$ or $u_i\rightarrow  +\infty$ as
$i\rightarrow \infty$. Consider, first, the case $u_i\rightarrow - \infty$.
Let $p' \in I^+(p)$ and let $q' \in I^-(q)$. Then there exists an integer
$\bar i$ such that for all $i > \bar i$ we have $p_i \in I^-(p')$ and
$q_i \in I^+(q')$. However, we have $q' \in M=I^+[M_{end}]$, so there
exists $r\in M_{end}$ such that $r\in I^-(q')$ and, hence,
$r\in I^-(q_i)$ for all $i>\bar i$. Applying the isometry $\chi_{-u_i}$ to
this statement, we find that $\chi_{-u_i}(r) \in I^-(p_i)$ and, hence,
$\chi_{-u_i}(r) \in I^-(p')$ for all $i>\bar i$. Since the Killing
orbits are future oriented in $M_{end}$, this implies that
$I^-(p')$ contains the entire Killing orbit through $r$, which, in turn,
implies that $I^-(p')\supset M_{end}$. However,
as in the proof of Lemma 3.1, this contradicts the fact
that, on account of global hyperbolicity, $J^-(p')\cap \Sigma$ must be
compact.

\parindent 20 pt

The proof for the case $u_i\rightarrow  +\infty$ follows by interchanging
the roles of $p$ and $q$ in the above argument. \sq

\bigskip
\parindent 0 pt
{\bf 4. Global extensions}
\parindent 20 pt
\medskip

The global extension of the class of black hole spacetimes specified
at the end of section 2 will be
based on the  use   of  so-called   Eddington-Finkelstein--type (EF)
coordinates. These coordinates were introduced in [6]
in a neighborhood of the
Killing horizon, $\tn$.
We begin by reviewing the construction of these
EF coordinates.

Let $\s$ be global  cross-section  of  ${\cal  N}$  obtained by
intersecting ${\cal  N}$ with a
Cauchy surface ${\cal C}$ as above.
Consider, now,  an open  subset, ${\~\s}$, of $\s$  on which local
coordinates $\x12$ can  be
introduced.   Denote  by  $\~{\tn}$  and  $\~{\ttn}$ the portions of ${\cal
N}$  and  ${\ttn}$, respectively,  generated by
null  geodesics  which  intersect  $\~  \s$.  First, we extend the
coordinates $\x12$ on $\~ \s$  to $\~{\ttn}$ by keeping  them constant
on the  null geodesic  generators of  $\~{\ttn}$.   Consider, now, the
unique  null  vector  field,  $\eta^a$,  on  $\~  \s$   satisfying
$\eta^a\xi_a=1$ and $\eta^aX_a=0$ for all $X^a$ which are  tangent
to $\~  \s$.  We define  the function  $r$ on  $\~{\ttn}$ to be the value of
the  affine  parameter  along  the null geodesic
generators of $\~{\ttn}$ starting on $\~ \s$ (with $r\vert_{\~ \s}=0$) with
tangent $\eta^a$.   Then $(r,\x12)$  comprise coordinates  on $\~{\ttn}$.
We  extend
the functions, $r,\x12$, from $\~{\ttn}$ to
${\cal O}_{\~{\tn}} \equiv \chi\{\~{\ttn}\}$ by  requiring
their values to be constant  along the Killing orbits through  $\~{\ttn}$.
Next  we  define  the  function  $u$  on ${\cal O}_{\~{\tn}}$ by the
conditions  $\xi^a\nabla_a  u  =1$  and  $u=0$  on  $\~{\ttn}$.  Then
$(u,r,\x12)$ yields  the desired
EF coordinate  system on  ${\cal O}_{\~{\tn}}$.

It follows [6] that the spacetime metric in ${\cal  O}_{\~{\tn}}$
takes   the   form
$$ds^2=-Fdu^2+2   du   dr+2  g_{u\a}du dx^\a+
g_{\a\b}dx^\a dx^\b,\eqno(12)$$
where
$F=-\xi^a\xi_a,g_{u\a}=\xi_a({\d \over\d x^\a})^a$, and $g_{\a\b}$
are smooth functions of $r,\x12$ in ${\to}_{\~{\tn}}$ such that
$F$ and $g_{u\a}$ vanish on ${\~\s}$ and $g_{\a\b}$ is  a
positive definite 2$\times$2-matrix.  (The Greek indices take  the
values $1,2$.) As shown in [6],
if the surface gravity, $\k$, of $\~{\tn}$ is constant and
nonvanishing, then
Kruskal--type coordinates  $(U,V,\x12)$ can be
defined in an open \ne, $\tilde {\cal O}\subset{\to}_{\~{\tn}}$ of $\~ {\tn}$
-- comprised by the
points   ${\cal    O}_{\~{\tn}}$   for    which
$\vert    UV\vert<\tilde \varepsilon(\x12)$ for some
suitably chosen positive function $\tilde \varepsilon(\x12)$ -- by
$$U = e^{\k u},\eqno(13)$$
$$V = - r e^{- \k u} {\rm exp}\bigl[ 2 \k \int^r_0 g(r', x^1, x^2) dr' \bigr],
\eqno(14)$$
where
$$g \equiv {1 \over F} - {1 \over 2r\k}.\eqno(15)$$
In the Kruskal coordinates,
the  spacetime  metric,  $g_{ab}$,  in  $\tilde {\cal O}$ takes
the form [6]
$$\eqalign{ds^2   =GdUdV+VH_\a  dUdx^\a+g_{\a\b}dx^\a
dx^\b,\cr}\eqno(16)$$
Here $G$, $H_\a$, and $g_{\a\b}$ are  smooth
functions  of  the  three  quantities  $UV,\x12$. The
precise range of the Kruskal type coordinates
in $\tilde  {\cal  O}$ is given by the inequality
$\vert    UV\vert<\tilde \varepsilon(\x12)$
together with the restrictions on the original
range of the coordinates $(\x12)$, and the inequality $U>0$.
Note that, by construction, $\tilde  {\cal  O}$ is comprised
by complete Killing orbits.

As in [6], we can construct a {\it local} extension of the spacetime to one
which contains a bifurcate Killing horizon as follows:
Let $\{\tilde \s_{_{(i)}}\}$ be a family of
charts covering $\s={\tn}\cap{\cal C}$. Since $\s$ is compact, we may take
$\{\s_{_{(i)}}\}$ be a finite collection of charts. For each $i$,
we introduce EF coordinates as above, and let
$(\tilde M_{_{(i)}},{{\tilde g}_{_{(i)}}}{}_{ab})$
be the spacetime defined by eq. (16), with
the range of the coordinates given by the original
range of the EF coordinates $(\x12)$ together with the inequality
$\vert    UV\vert<\tilde \varepsilon_{_{(i)}}(\x12)$.
In other words, the spacetime
manifold is $\tilde M_{_{(i)}}=\tilde\sigma_{_{(i)}}\times$\re$^2$ with metric
given by (16), but with
the restriction $U > 0$ now dropped. As in [6], we may ``patch together"
the spacetimes $(\tilde M_{_{(i)}},\tilde g_{_{(i)}}{}_{ab})$
to obtain a new spacetime $(\~ M, \tilde g_{ab})$
by taking the union of the $(\tilde M_{_{(i)}},\tilde {g}_{_{(i)}}{}_{ab})$
and then identifying the points labeled by
$(U_{_{(i)}}, V_{_{(i)}}, x^1_{_{(i)}}, x^2_{_{(i)}})$
and $(U_{_{(j)}}, V_{_{(j)}}, x^1_{_{(j)}}, x^2_{_{(j)}})$ if and only if
$(x^1_{_{(i)}}, x^2_{_{(i)}})$ corresponds to the same point of $\s$ as
$(x^1_{_{(j)}}, x^2_{_{(j)}})$ and $U_{_{(i)}} = U_{_{(j)}}$, $V_{_{(i)}}
= V_{_{(j)}}$ (see the bottom of P. 2651 of [6] for details).
Note that the functions $U$ and $V$
are then globally well defined in $(\~ M, \tilde{g}_{ab})$.
It may be verified straightforwardly
(see [6]) that $(\~ M, \tilde{g}_{ab})$ is an extension of the
spacetime $({\tu}, g_{ab})$, where
${\tu}=\bigcup_{_{i}} \tilde  {\cal  O}_{_{(i)}} \subset M$.
Here the isometric
embedding $\phi:{\tu}\ra \~ {M}$ arises from the maps
$\~\phi_{_{(i)}}:\tilde {\cal  O}_{_{(i)}}\ra{\~ M_{_{(i)}}}$ which
take each point  of $\tilde  {\cal  O}_{_{(i)}}$
into the point in $\~ M_{_{(i)}}$ having the same values of the coordinates
$(U_{_{(i)}}, V_{_{(i)}}, x^1_{_{(i)}}, x^2_{_{(i)}})$.
It also is  straightforward  to
verify that the hypersurfaces defined by $U=0$ and $V=0$  comprise
a  bifurcate  Killing  horizon,  $\cal H$,  in  $\~ M $ and that the
spacetime $(\~ M, \tilde{g}_{ab})$ possesses a reflection isometry
defined by $U \rightarrow -U$, $V \rightarrow -V$
about the bifurcation surface, $S$,
given by $U=V=0$. Finally, we note that the image
of ${\tn}$ under ${ \phi}$ comprises the portion of $\cal H$  defined
by $V=0$, $U>0$.

Note that, using the isometric imbedding,
$ \phi$, one can show that the Killing orbits generating
$\tn$ must be geodesically incomplete in the past rather than the
future; equivalently, the surface gravity, $\kappa$ defined by eq. (1)
must be positive rather than negative. Namely, suppose, on the contrary,
that the orbits on $\tn$ were future incomplete. Then the
hypersurfaces of constant $UV$ sufficiently close to $\tn$ would be
spacelike in $I^-[\tn]$, and, consequently, in this region $UV$
would have to decrease along all
future-directed causal curves. From this it would follow that for any
$p \in I^-[\tn]$ sufficiently close to $\tn$,
the image under $\phi$ of any future-directed null geodesic
starting at $p$ would remain in $\~ M$ at least until reaching
$\cal H$. However,
this would contradict the fact that some future-directed null geodesics
from $p$ must reach $M_{ext}$. Thus, the horizon,
$\tn$, of $(M,g_{ab})$ cannot be ``upside down" with respect to the
usual orientation of the horizon occurring, e.g., in the Schwarzschild
solution.

\medskip

The above local extension will now be ``globalized" with the help of
lemma 3.4 together with the following general lemma:

\medskip\parindent 0pt
\leftskip=3 truecm

\item{\it Lemma 4.1:} Let $(M, g_{ab})$ be an $n$-dimensional
spacetime without boundary, and let $(M', {g'}_{ab})$ be an
$n$-dimensional spacetime with boundary
$\partial M'$. Let ${\cal O}'$ be an $n$-dimensional
submanifold with boundary of $M'$,
such that $\partial M' = \partial {\cal O}'$.
Let ${\cal Q}$ be a {\it closed} subset of $M$ such that the differential
structure of $M$ induces
an $n$-dimensional manifold with boundary structure on ${\cal Q}$.
Suppose that there exists an isometry $\psi : {\cal O}' \ra {\cal Q}$.
Define ${\hat M} = (M \cup M') / \psi$, i.e., we identify points $x, x'$
in the union
of $M$ and $M'$ if and only if $x \in {\cal Q}$, $x' \in {\cal O}'$ and
$\psi (x') = x$.
Then ${\hat M}$ has the natural structure of a
(Hausdorff) manifold
without boundary and the spacetime $({\hat M}, {\hat g}_{ab})$
is an extension of $(M, g_{ab})$, where ${\hat g}_{ab}$ is the metric
on ${\hat M}$ naturally induced from $g_{ab}$ and ${g'}_{ab}$.

\leftskip=1 truecm

 {\it Proof:} Consider the charts on ${\hat M}$ which arise either from
charts on $M$ or charts on ${\rm int} (M')$. Then it is straightforward to
verify
that these charts comprise a compatible family which cover ${\hat M}$,
thereby giving ${\hat M}$ the structure of a manifold without boundary.
The Hausdorff property is obvious except for pairs of points in ${\hat M}$
arising from points $(x', x)$
with $x'$ in the closure of ${\cal O}'$ in $M'$ and $x$ in the closure of
${\cal Q}$
in $M$. However, since by hypothesis ${\cal Q}$ is closed in $M$, $x$ is
the image under $\psi$ of a point in $M'$, from which it follows immediately
that $x$ can be Hausdorff separated from $x'$.
It is straightforward to verify that $({\hat M}, {\hat g}_{ab})$
is an extension of $(M, g_{ab})$. \sq

\medskip\parindent 20pt

We now state and prove our main result:

\medskip\parindent 0pt
\leftskip=3.1 truecm

\item{\it Theorem 4.2:} Let $(M,g_{ab})$ be a spacetime of the class
specified at the end of section 2. Suppose, further, that surface gravity,
$\k$, is a non-zero constant on the horizon $\tn$. Then $(M,g_{ab})$ can be
extended {\it globally} so that in the enlarged spacetime the image of
$\tn$ will be a proper subset of a bifurcate Killing horizon.
Furthermore, an extension $(M^*, g^*_{ab})$
can be chosen so that the original one-parameter group of isometries
$\chi_u$ extends to $(M^*, g^*_{ab})$, and such that $(M^*, g^*_{ab})$
possesses a ``wedge reflection" isometry, $w$,
about the bifurcation surface $S$.

\leftskip=1 truecm

{\it Proof:} Let $(\~ M,\tilde{g}_{ab})$ be the local extension of $(M,g_{ab})$
constructed above. Let $\varepsilon$ be any positive function
on $\s$ such that at each $(\x12)$, we have
$\varepsilon(\x12) < \tilde \varepsilon_{_{(i)}}(\x12)$ for some $i$.
Let $M'$ be the submanifold with boundary of $\~ M$ consisting of the
points of $\~ M$ which satisfy
$$\vert    UV\vert \cases{\leq \varepsilon(\x12),&if
$U > 0$;\cr   <\varepsilon(\x12),&if $U \leq 0$.\cr}\eqno(17)$$
Let ${\cal O}'$ be the
subset of $M'$ satisfying $U>0$. Let ${\cal Q}$ be the pre-image of
${\cal O}'$ under the isometric embedding map $\phi$ provided by the
above local extension. It follows directly from lemma 3.4 that
${\cal Q}$ is a closed subset of $M$. Hence, the conditions of lemma 4.1
hold. Thus, we obtain a global extension,
$({\hat M}, {\hat g}_{ab})$,  of $(M, g_{ab})$ by ``patching it"
to $(M', g'_{ab})$ in the manner explained in lemma 4.1.

\parindent 20pt

Now, consider the new spacetime $({\hat M'}, {\hat g'}_{ab})$ obtained
from $({\hat M}, {\hat g}_{ab})$
by reversing its time orientation and also
eliminating the points which came from points in
$(\~ M, \tilde{g}_{ab})$ satisfying both $U < 0$ and
$\vert    UV\vert > \varepsilon(\x12)/2$.
Introduce new coordinates $(U', V')$ in a neighborhood
of the bifurcate Killing horizon of $({\hat M'}, {\hat g'}_{ab})$
by $U' = - U$, $V' = - V$.
Since $(\~ M, \tilde{g}_{ab})$ possesses the reflection
isometry $U \rightarrow -U$,
$V \rightarrow -V$ about the bifurcation surface, it is clear that the
construction of the previous paragraph now can be repeated, with
$(M', g'_{ab})$ replaced by $({\hat M'}, {\hat g'}_{ab})$. The isometries
$\chi_u$ clearly extend to the resulting
extension, which we denote as $(M^*, g^*_{ab})$,
and $(M^*, g^*_{ab})$ manifestly also possesses a ``wedge reflection"
isometry about the bifurcation surface $U = V = 0$. \sq

\medskip

We conclude this section with the following proposition,
which shows that the extension $(M^*, g^*_{ab})$ satisfies all of the
properties needed for the
application of the results of [8].

\medskip\parindent 0pt
\leftskip=3.6 truecm

\item{\it Proposition 4.3:} Let $(M^*, g^*_{ab})$ be the
``wedge reflection" invariant extension of
$(M, g_{ab})$ constructed in theorem 4.2. Let
${\cal D} = M \setminus {\cal B}$ denote the domain
of outer communications of $(M, g_{ab})$. Then there exists a globally
hyperbolic open region ${\cal V} \subset M^*$
which is invariant under the
isometries $\chi_u$ such that $\bar{\cal D} \subset {\cal V}$, where
$\bar{\cal D}$ denotes the closure of ${\cal D}$ in $M^*$.

\leftskip=1 truecm

 {\it Proof:} Let ${\cal C}$ be a smooth, spacelike Cauchy surface for
$(M, g_{ab})$. Let $\tilde{{\cal C}}$ be the hypersurface in $(M, g_{ab})$
obtained by removing the portion of ${\cal C}$ lying in ${\cal B}$ and
replacing it with the portion of $\tn$ lying in the causal past of
$\s = \tn \cap {\cal C}$. Finally, let ${\cal C}^*$ be the hypersurface
in $(M^*, g^*_{ab})$ obtained by adjoining to $\tilde{{\cal C}}$
the bifurcation surface, $S$,
together with the image of $\tilde{{\cal C}}$ under the wedge reflection
isometry, $w$. Then ${\cal C}^*$ is a closed, achronal, edgeless set. Let
$p \in {\cal D}$. If $p \in J^+[{\cal C}]$, then any past inextendible
causal curve through $p$ intersects ${\cal C}$ outside of ${\cal B}$,
and, hence, must intersect ${\cal C}^*$. On the other hand, if
$p \in J^-[{\cal C}]$, then any future-inextendible causal curve $\gamma$
must intersect ${\cal C}$. If
$\gamma$  intersects ${\cal C}$ in ${\cal D}$, then
clearly it intersects ${\cal C}^*$. But if $\gamma$ intersects ${\cal C}$ in
${\cal B}$, then $\gamma$ must cross $\tn$ to the causal past of $\s$, and,
hence, $\gamma$ also must intersect ${\cal C}^*$. Since ${\cal D}$ is open,
this establishes that ${\cal D} \subset {\cal V}$, where
${\cal V} \equiv {\rm int} D[{\cal C}^*]$ and $D$ denotes the domain of
dependence. Thus, in order to show that $\bar{\cal D} \subset {\cal V}$,
it suffices to show that there cannot exist a point
$p \in \partial{\cal D}$ such that
$p \in \partial{\cal V} = \partial D[{\cal C}^*]$. However, if
$p \in \partial{\cal D}$, then $p$ must lie either on the
portion, ${\cal H}^+$, defined by $V = 0, U \geq 0$ or the portion,
${\cal H}^-$, defined by $U = 0, V \leq 0$, of
the bifurcate Killing horizon of $(M^*, g^*_{ab})$. In the former case,
we have $p \in J^+[{\cal C}^*]$, so if $p \in \partial D[{\cal C}^*]$, we must
have $p \in H^+[{\cal C}^*]$, where $H^+$ denotes the future
Cauchy horizon. However, in that case, $p$ must lie on a past inextendible
null geodesic which does not intersect ${\cal C}^*$.
But if $p \in {\cal H}^+$, then either $p \in {\cal C}^*$ -- in which case
every null geodesic through $p$ clear intersects ${\cal C}^*$ --
or $p \in I^+[{\cal C}^*]$. In the latter case, every past inextendible
causal curve through $p$ either enters ${\cal D}$ to the future
of ${\cal C}^*$ -- in which
case, by the above argument, it intersects ${\cal C}^*$ -- or it coincides
with the null geodesic generator of the horizon -- in which case it
also intersects
${\cal C}^*$. The proof for the case $p \in {\cal H}^-$ proceeds similarly.

\parindent 20 pt

Finally, we note that $D[{\cal C}^*]$ may be characterized by the fact that
$q \in D[{\cal C}^*]$ if and only if every inextendible causal curve through
$q$ enters either $\bar{\cal D}$ or its image under the wedge reflection
isometry, $w$. However, since $\bar{\cal D}$
is invariant under $\chi_u$, it follows that $D[{\cal C}^*]$ and, hence,
${\cal V} = {\rm int} D[{\cal C}^*]$ also is invariant under $\chi_u$. \sq

\bigskip
\parindent 0 pt
{\bf 5. Extensions of matter fields}
\parindent 20 pt
\medskip

In this section, we analyze the extendibility of ``matter fields" defined
on a spacetime $(M, g_{ab})$ in the class we are considering. More
precisely, let ${T^{a_1...a_k}}_{b_1...b_l}$ be an
arbitrary, smooth tensor field of type $(k,l)$
on $(M, g_{ab})$ which is invariant
under the isometries $\chi_u$, so that
$${\L}_\xi {T^{a_1...a_k}}_{b_1...b_l} = 0.\eqno(18)$$
Under what conditions can ${T^{a_1...a_k}}_{b_1...b_l}$ be extended
to a smooth, isometry invariant tensor field on the extension,
$(M^*, g^*_{ab})$ constructed in theorem 4.2? For
a tensor field of type $(0,0)$ (i.e., a function), it is straightforward to
verify
that isometry invariance on $M$ implies extendibility to $M^*$.
However, this result does not hold for tensor fields of higher type.
For example, in a region covered by EF coordinates
the one-form $du$ is smooth and isometry invariant, but cannot be
smoothly extended to $M^*$, as can be seen from the fact that in
terms of the Kruskal coordinates, we have $du = dU/U$.

Since the spacetime metric extends smoothly to $(M^*, g^*_{ab})$,
without loss of generality it suffices to consider the case of tensor
fields of type $(0,l)$. In the following, we shall supress indices and
denote such a tensor field by $T$.
The only obstruction to a smooth extension of $T$
will arise from the behavior of $T$ near the bifurcate Killing horizon,
so without any loss of generality,
we may restrict attention to the region of $(M^*, g^*_{ab})$ where
the Kruskal coordinates $(U, V)$ are well defined. Without loss of
generality, we may also restrict
attention to a subregion covered by a single patch
of the coordinates $(\x12)$.
Let ${\cal O}$ denote such a subregion.
In addition, it will be useful,
initially, to further restrict attention to the (``right wedge")
portion, ${\cal R}$, of ${\cal O}$
consisting of points satisfying $U>0, V<0$. Thus,
${\cal R} \subset {\cal D}$, where ${\cal D}$ denotes the domain of
outer communications of $M$.
Our first lemma is the following:

\medskip\parindent 0pt
\leftskip=3 truecm

\item{\it Lemma 5.1:} Within ${\cal R}$, the one-forms
$dU/U$, $dV/V$, $dx^1$, and $dx^2$ are invariant under the isometries
$\chi_u$ and are linearly independent at each point.

\leftskip=1 truecm

 {\it Proof:} The linear independence of $dU/U$, $dV/V$, $dx^1$,
and $dx^2$ in ${\cal R}$
follows immediately from the linear independence of the
Kruskal coordinate basis $dU$, $dV$, $dx^1$,
and $dx^2$. Isometry invariance of $dx^1$,
and $dx^2$ follows immediately from the isometry invariance of
the EF coordinates $(\x12)$. Isometry invariance of $dU/U$ follows
from the fact that the EF coordinate basis element $du$ is isometry
invariant, and from the
definition (13) of $U$, we have $dU/U = du$. Finally, the definitions
of $U$ and $V$ immediately imply that $UV$ can be expressed as a
function of $(r, \x12)$, and, hence is isometry invariant. Hence,
$d(UV)/(UV) = dU/U + dV/V$ is isometry invariant, which, in turn,
implies that $dV/V$ is isometry invariant. \sq

\medskip\parindent 20pt

Given a tensor field $T$ of type $(0,l)$ in ${\cal R}$, we expand it
in the basis constructed from $dU/U$, $dV/V$, $dx^1$, and $dx^2$.
Since this basis is isometry invariant, the tensor field $T$ will be
isometry invariant if and only if all of the functions appearing in
the basis expansion
are themselves isometry invariant. This, in turn, is equivalent to
requiring that these basis expansion
functions can be expressed as functions of $(UV, \x12)$
only. Thus, in ${\cal R}$, the basis expansion of any isometry invariant
tensor field, $T$, of type $(0,l)$ takes the schematic form
$$T = \sum f^{(p),(q),(r),(s)} (UV, \x12)
(dU/U)^{(p)} (dV/V)^{(q)} (dx^1)^{(r)} (dx^2)^{(s)},\eqno(19)$$
where the superscripts $p,q,r,s$ (with $p+q+r+s = l$)
denote the number of times the basis element appears in the particular
term in the expansion. Since $dU$, $dV$, $dx^1$,
and $dx^2$ remain smooth and linearly independent in the extended
spacetime, it follows that $T$ is extendible if and only if each term in
the above basis expansion is separately extendible. The next
lemma then follows immediately by inspection of the form of this
basis expansion

\medskip\parindent 0pt
\leftskip=3 truecm

\item{\it Lemma 5.2:} Let $T$ be a smooth, isometry invariant
tensor field in ${\cal R}$. Then $T$ is smoothly extendible to the
hypersurface $V = 0, U>0$ in $M$ (i.e., to the original Killing horizon
$\tn$) if and only if each basis expansion function
$f^{(p),(q),(r),(s)}$ can be written as
$$f^{(p),(q),(r),(s)} (UV, \x12) = (UV)^q \alpha^{(p),(q),(r),(s)} (UV, \x12),
\eqno(20)$$ where $\alpha^{(p),(q),(r),(s)}$ is a smooth function of its
arguments. Similarly, $T$ is smoothly extendible to the
hypersurface $U = 0, V < 0$ in $M^*$ if and only if each
$f^{(p),(q),(r),(s)}$ can be written as
$$f^{(p),(q),(r),(s)} (UV, \x12) = (UV)^p \beta^{(p),(q),(r),(s)} (UV, \x12),
\eqno(21)$$
where $\beta^{(p),(q),(r),(s)}$ is smooth. Finally, $T$ is smoothly extendible
as an isometry invariant tensor field
to a neighborhood of the entire bifurcate Killing horizon $UV = 0$
(and, hence, to all of $M^*$) if and only if
$$f^{(p),(q),(r),(s)} (UV, \x12) =
(UV)^{{\rm max}(p,q)} \gamma^{(p),(q),(r),(s)} (UV, \x12),\eqno(22)$$
where $\gamma^{(p),(q),(r),(s)}$ is smooth.

\leftskip=1 truecm
\medskip\parindent 20pt

If $T$ is given as a smooth tensor field on $(M, g_{ab})$, then we know
that as a tensor field on ${\cal R}$
it is smoothly extendible to $\tn$. Thus, each $f^{(p),(q),(r),(s)}$
must take the form
$$f^{(p),(q),(r),(s)} (UV, \x12) = (UV)^q \alpha^{(p),(q),(r),(s)} (UV, \x12).
\eqno(23)$$
It then follows immediately from lemma 5.2 that the necessary and
sufficient condition for the extendibility of $T$ to an isometry invariant
tensor field on $(M^*, g^*_{ab})$ is that for all terms in the basis
expansion with $p > q$ we have
 $$\alpha^{(p),(q),(r),(s)} (UV, \x12) = (UV)^{p-q}
{\tilde \alpha}^{(p),(q),(r),(s)} (UV, \x12),\eqno(24)$$
where ${\tilde \alpha}^{(p),(q),(r),(s)}$ is smooth.

Our final result is the following:

\medskip\parindent 0pt
\leftskip=3.1 truecm

\item{\it Theorem 5.3:} Let $(M, g_{ab})$ be a spacetime in the class
specified at the end of section 2, and suppose, furthermore, that
$(M, g_{ab})$ either is static or is stationary-axisymmetric with a
``$t - \phi$" reflection isometry. Let $i$ denote, respectively, the
``$t$" or ``$t - \phi$" reflection isometry
in ${\cal D}$. Let $T$ be a smooth tensor field
on $(M, g_{ab})$ which is invariant under the isometries $\chi_u$
and also is invariant under $i$ in ${\cal D}$. Then $T$ is smoothly
extendible to an isometry invariant tensor field on $(M^*, g^*_{ab})$.

\leftskip=1 truecm

 {\it Proof:} Consider the restriction of $T$ to the region,
${\cal R}$, of $(M, g_{ab})$ defined above. Since $T$ is given as
smooth on $(M, g_{ab})$, it obviously is smoothly extendible to
$\tn$ given by $V=0, U >0$. However, the isometry $i$ extends to
the boundary of ${\cal R}$ in $(M^*, g^*_{ab})$ in such a way as to
map the hypersurface $V=0, U >0$ into the hypersurface
$U=0, V < 0$. Since
$T$ is invariant under $i$, it follows that $T$ must also be extendible
to the hypersurface $U=0, V < 0$. It then follows from lemma 5.2 that
$T$ is extendible to $(M^*, g^*_{ab})$. \sq

\vfill\eject

\medskip\parindent 0pt
{\bf Acknowledgements}
\medskip\parindent 20pt

This research was supported in part by NSF grant PHY 92-20644, OTKA
grants F14196, T016246.
One of us (IR) wishes to thank to the Fermi Institute for its hospitality
during a visit, which was made possible by a supplement to grant PHY 92-20644
provided by the International Division of the NSF and MTA (Hungarian
Academy of Sciences) grant 77-94 for scientific collaborations.

\leftskip 1.1truecm
\medskip\parindent 0pt
{\bf References}
\medskip
\parindent 0pt

\item{[1]} R.M. Wald : {\it General Relativity} (University of Chicago
Press, 1984)

\item{[2]} S.W. Hawking: Commun. Math. Phys. {\bf 25}, 152-166 (1972)

\item{[3]} S.W. Hawking and G.R.F. Ellis: {\it Large Scale Structure
of Spacetime} (Cambridge University Press, Cambridge, 1973)

\item{[4]} B. Carter: Phys. Rev. Lett. {\bf 26}, 331-333 (1971)

\item{[5]} B. Carter: in {\it Black holes}, ed. C. DeWitt and B.S. Dewitt
(New York: Gordon \& Breach, 1973)

\item{[6]} I. R\'acz  and R.M. Wald:  Class.  Quant.  Grav.  {\bf
9}, 2643-2656 (1992)

\item{[7]} I. R\'acz: J. Math Phys. {\bf 34}, 2448-2464 (1993)

\item{[8]} B.S. Kay and R.M. Wald: Phys.  Rep. {\bf 207},  49-136
(1991)

\item{[9]} M. Heusler: {\it Black Hole Uniqueness Theorems}, Cambridge
Lecture Notes in Physics (Cambridge, in press)

\item{[10]} P.T. Chru\' sciel and R.M. Wald: Common. Math. Phys. {\bf 163},
561-604 (1994)

\vfill\eject \end